\documentclass[12pt]{article}
\usepackage{amssymb}
\usepackage{amsthm}
\usepackage{eucal}
\theoremstyle{plain}
\newtheorem{theorem}{Theorem}[section]

\newtheorem*{m-p}{Rule of Inference}

\begin{document}
\voffset=-20pt
\hoffset=-60pt
\textwidth=180mm
\textheight=210mm

\markright{\it International Journal of Theoretical Physics\/\rm,
\bf 42\sl, No.~12 (2003)}

\vbox to 2cm{\vfill}

\begin{flushleft}
{\LARGE\bf Equivalences, Identities, Symmetric\\
Differences, and Congruences\\
in Orthomodular Lattices\\}

\end{flushleft}

\bigskip
\bigskip
\bigskip
\parindent=2cm\hangindent=4cm\baselineskip=15pt

\bf Norman D.~Megill$\,$\footnote{Boston Information Group,
Belmont MA 02478, U.~S.~A.; E-mail: nm@alum.mit.edu;
Web page: http://www.metamath.org}
and
Mladen Pavi\v ci\'c$\,$\footnote{University of Zagreb,
Gradjevinski fakultet, Ka\v ci\'ceva 26, HR-10001 Zagreb, Croatia;\break
E-mail: mpavicic@irb.hr; Web page: http://m3k.grad.hr/pavicic}

\rm
\medskip
\vrule height .3ex width 11.70493cm depth -.05ex

\parindent=2cm\hangindent=2cm\baselineskip=15pt\medskip
It is shown that operations of equivalence cannot serve for
building algebras which would induce orthomodular lattices
as the operations of implication can.
Several properties of equivalence operations have been
investigated. Distributivity of equivalence terms and several
other 3 variable expressions involving equivalence terms have
been proved to hold in any orthomodular lattice. Symmetric
differences have been shown to reduce to complements of
equivalence terms. Some congruence relations related to 
equivalence operations and symmetric differences have been 
considered. 
\medskip

\vrule height .3ex width 11.71043cm depth -.05ex

\smallskip


\smallskip


\parindent=4.1cm
quantum identity, equivalence relation,

\parindent=4.1cm
congruence relation, symmetric difference.

\parindent=2cm
\smallskip

\vrule height .3ex width 11.71043cm depth -.05ex

\vbox to 1cm{\vfill}

\vfill\eject
\markright{\rm N.~Megill and M.~Pavi\v ci\'c, \
   \it Quantum Equivalences and Congruences}

\parindent=20pt\hangindent=0pt
\vbox to 3mm{\vfill}
\baselineskip=17pt
\section{\large INTRODUCTION}
\label{sec:intro}

It is well-known that any orthomodular lattice equations and
conditions generated with at most two generators has classical and
quantum constants, variables, and operations---altogether
96 so-called Beran expressions. \cite{beran,mpqo01,mpqo02}
All quantum constants, variables, and operations are fivefold
defined by means of classical ones. \cite{mpcommp99}
Also, whenever all classical constants in an orthomodular lattice
commute the orthomodular lattice becomes the Boolean algebra and
quantum constants, variables, and operations reduce to classical
ones.

Classical constants are 0 and 1 (Beran expressions 1 and 96),
classical variables are $a,b$ (22,39) and their complements
$a^\perp,b^\perp$ (58,75). Quantum constants are: quantum 0's
(17,33,49,65,81) and quantum 1's (16,32,48,64,80). Quantum
variables are: quantum $a$ (6,38,54,70,86), $b$ (7,23,55,71,87),
$a^\perp$ (11,27,43,59,91), and $b^\perp$ (10,26,42,74,90).
\cite{mpqo01,mpqo02}

In this paper we show that the binary operations in an
orthomodular lattice can be divided into two groups.
A group containing operations which together with
complementation can be used to express any other operation, and
a group which does not enable this. To the former group belong
joins and meets and to the latter operations of equivalence.
Both of them again have classical and quantum representatives.
Classical meet, $\cup$ and join, $\cap$ with $a,b,a^\perp,a^\perp$
(Beran expressions 2--5 and 92--95) and quantum meets and joins
(12--21, 28--37, 44--53, 60--69, and 76--85) from the former
group and  classical equivalence, $\equiv$ and its complement
(88 and 9) and quantum  equivalences, $\equiv_i$, $i=1,\dots,5$ and
their complements (24,25,40,41,56,57,72,73,8,89). \cite{mpqo01,mpqo02}
In the field of quantum logic and orthomodular lattices
meet and joins have, in the literature, been given various other names
depending on the distribution of complements. E.g., implication,
conditional, projection, skew operations \cite{beran}, sharp and flat
operations \cite{pykacz-notre-d,hoog-pyk}, etc. Also operations of
equivalence and their complements have been given other names like
(symmetric) (classical and quantum) identity \cite{pav89,pav93,p98}
and asymmetric (quantum) identities \cite{mpcommp99,mpqo01,mpqo02}
and symmetric difference \cite{beran,dorfer-lang} and non-commutative
symmetric differences \cite{dorfer}.

In this paper we will concentrate on the equivalence operations
and we first show that one cannot use equivalence operations to
express other operations and therefore that an equivalence algebra
cannot induce orthomodular lattices in a way the implication
algebras can. (Cf.~implication algebras given by
\cite{kimble69,clark73,piziak74,abbott76,abbott,gaca80,harde81a,harde81b,mpijtp98,mpijtp03})
In Sec.~\ref{sec:open} we give a solutions to
previously open 3 variable problems of expressions containing
symmetric equivalence terms. In the end, in Sec.~\ref{sec:s-diff} we
prove that recently introduced non-commutative symmetric differences
\cite{dorfer-lang} are nothing but complements of asymmetric equivalence
relations and that therefore the majority of the  results obtained in
\cite{dorfer-lang} directly follow from the results previously obtained in
\cite{mpcommp99,mpqo01,mpqo02}.

\bigskip
\section{\large NO-GO FOR EQUIVALENCE ALGEBRA}
\label{sec:equiv-alg}

All implications from a quantum logic (an orthomodular lattice) reduce
to the classical one in a classical theory (a Boolean algebra).
So, as we show in \cite{mphpa98}, not only $a\leftrightarrow_i b$ but
also $(a\to_i b)\cap(b\to_j a)$, $i\ne j$
($i=0,1,\dots,5$, where $\to_i$ correspond to Beran expressions:
94,78,46,30,62,14, respectively) must reduce to $a\leftrightarrow_0 b$
in a classical theory. To handle the Beran expressions we use
programs \tt beran \rm and \tt bercomb\rm. \cite{mpqo01} Let us have
a look at what we get in an orthomodular lattice in Table 1, where
B($a,b$) are Beran expressions (5 of 96 ones given in \cite{beran}).

{\parindent=5pt
\begin{tabular}{|lr||c|c|c|c|c|c|} \multicolumn{8}{c}{}\\ \hline
\ $^i _\downarrow\ ${\huge$\backslash$}&$^j_\rightarrow$
& $b\rightarrow_0 a$ & $b\rightarrow_1 a$ & $b\rightarrow_2 a$ &
$b\rightarrow_3 a$ & $b\rightarrow_4 a$ & $b\rightarrow_5 a$
\\ \hline \hline
\multicolumn{2}{|c||}{$a\rightarrow_0b$} & B$_{88}$(a,b)
   & B$_{56}$(a,b)  &  B$_{24}$(a,b)  &
   B$_{40}$(a,b) &  B$_{72}$(a,b)
   & B$_{8}$(a,b)\\ \hline
\multicolumn{2}{|c||}{$a\rightarrow_1b$} & B$_{72}$(a,b) & B$_{8}$(a,b)
& B$_{8}$(a,b) &
B$_{8}$(a,b) & B$_{72}$(a,b) & B$_{8}$(a,b)\\ \hline
\multicolumn{2}{|c||}{$a\rightarrow_2b$} & B$_{40}$(a,b) &
B$_{8}$(a,b) & B$_{8}$(a,b) &
B$_{40}$(a,b) & B$_{8}$(a,b) & B$_{8}$(a,b)\\ \hline
\multicolumn{2}{|c||}{$a\rightarrow_3b$} & B$_{24}$(a,b) &
B$_{8}$(a,b) & B$_{24}$(a,b) &
B$_{8}$(a,b) & B$_{8}$(a,b) & B$_{8}$(a,b)\\ \hline
\multicolumn{2}{|c||}{$a\rightarrow_4b$} & B$_{56}$(a,b) &
B$_{56}$(a,b) & B$_{8}$(a,b) &
B$_{8}$(a,b) & B$_{8}$(a,b) & B$_{8}$(a,b)\\ \hline
\multicolumn{2}{|c||}{$a\rightarrow_5b$} & B$_{8}$(a,b) &
B$_{8}$(a,b) & B$_{8}$(a,b) &
B$_{8}$(a,b) & B$_{8}$(a,b) & B$_{8}$(a,b)\\ \hline
\multicolumn{8}{c}{}\\
\multicolumn{8}{c}{Table 1: Products $(a\rightarrow_i b)
\cap(b\rightarrow_j a)$, $i=0,\dots,5$, $j=0,\dots,5$}\\
\multicolumn{8}{c}{reduced to Beran expressions.} \\
\multicolumn{8}{c}{}\\
\end{tabular}
}

The expressions B($i$), $i=24,40,56,72$ are asymmetrical
and at first we would think it would be inappropriate to name
them equivalence operations. But we were able to prove
Theorem \ref{th:other-eq} and Theorem \ref{th:boole-eq}
below and therefore we define symmetric and asymmetric
equivalence operations as follows:
\begin{eqnarray}
a\equiv_0 b\ &{\buildrel\rm def\over =}&\ (a'\cup b)\cap
(a\cup b')\ \quad\qquad\qquad(={\rm B}_{88}(a,b))\nonumber\\
a\equiv_1 b\ &{\buildrel\rm def\over =}&\ (a\cup b')\cap
(a'\cup (a\cap b))\quad\qquad(={\rm B}_{72}(a,b))\nonumber\\
a\equiv_2 b\ &{\buildrel\rm def\over =}&\ (a\cup b')\cap
(b\cup (a'\cap b'))\ \ \,\qquad(={\rm B}_{40}(a,b))\nonumber\\
a\equiv_3 b\ &{\buildrel\rm def\over =}&\ (a'\cup b)\cap
(a\cup (a'\cap b'))\ \ \qquad(={\rm B}_{24}(a,b))\nonumber\\
a\equiv_4 b\ &{\buildrel\rm def\over =}&\
(a'\cup b)\cap(b'\cup (a\cap b))\ \ \ \qquad
(={\rm B}_{56}(a,b))\nonumber\\
a\equiv_5 b\ &{\buildrel\rm def\over =}&\
(a\cup b)\cap(b'\cup a')\qquad\qquad\quad\
(={\rm B}_{8}(a,b)).\nonumber
\end{eqnarray}

\begin{theorem}\label{th:other-eq}
{\rm[\protect\cite{mpcommp99}]} Ortholattices in which
\begin{eqnarray}
a\equiv_i b=1\qquad \Leftrightarrow\qquad a=b, \qquad\qquad
i=1,\dots,5,\label{eq:qm-as-id}
\end{eqnarray}
hold are orthomodular lattices and vice versa.
\end{theorem}

\begin{theorem}\label{th:boole-eq}
{\rm[\protect\cite{p98}]} Ortholattices in which
\begin{eqnarray}
a\equiv_o b=1\qquad \Leftrightarrow\qquad a=b\label{eq:cl-as-id}
\end{eqnarray}
holds is a Boolean algebra and vice versa.
\end{theorem}

A natural question which springs from these theorems
is whether one can express joins and complements by means
of the two above-defined operations of equivalence, i.e.,
whether  ``equivalence algebras,'' analogous to implication
algebras \cite{abbott,abbott76,mpijtp98,mpijtp03}, can be
formulated. In \cite{p98} we answer such a question
for the symmetric equivalence, $\equiv_5$ in the negative.
By the following theorem we answer to this question
in the negative for the classical, $\equiv_0$ and the
asymmetric equivalences, $\equiv_i$, $i=1,\dots,4$ as well.
Therewith we also prove that an ``equivalence algebra''
cannot be formulated.

\medskip
\begin{theorem}\label{th:id-al} Orthocomplementation in
an orthomodular lattice can be expressed as $a'=
a\>\equiv_i\>0,\ \ i=0,1,\dots,5$. However, classical and
quantum joins (including implications) and classical and
quantum meets and their complements in an orthomodular
lattice cannot be expressed by means of the operations
of equivalence.
\end{theorem}

\begin{proof}Free orthomodular lattices with two generators
(expressions with two elements) can be represented by the
direct product M$_{\rm{O}2}\,\times\,2^4$ \cite{beran}. Denoting
the elements of the Boolean algebra $2^4$ by $b_1=(0,0,0,0)$,
$b_2=(1,0,0,0)$,\dots,$b_{16}=(1,1,1,1)$, we can write
down all 96 elements of the lattice in the form $(a_i,b_j),
i=1,\dots,6,j=1,\dots,16$, where $a_i$ are the elements of
the orthomodular lattice M$_{\rm{O}2}$ (also called OM$_6$;
Fig.~(1) of \cite{p98}). We can
easily check that $(a_i,b_{12})$ through $(a_i,b_{15}),i=1,\dots,6$,
are exactly all six joins (quantum and classical; among them,
of course of implications), while $(a_i,b_2)$ through $(a_i,b_5)$
are their negations, i.e., quantum and classical meets. When we
look at the Boolean part only we can see that they are all
characterised with an odd number of 1's (0's) (either one or
three).

Looking at the Boolean parts of the other Beran expressions
we find that they all have an even number of 1's and 0's.
Quantum and classical 0's are represented by (0,0,0,0),
1's by (1,1,1,1), x's by (1,1,0,0), -x's by (0,0,1,1),
y's by (1,0,1,0), -y's by (0,1,0,1), equivalences by
(1,0,0,1) and their negations by (0,1,1,0). Simple checking
then shows that whatever expression we introduce into
equivalences and/or their negations we always end up with
expressions whose Boolean parts have only even number of 1's
and 0's. This proves the theorem.
\end{proof}

\bigskip
\section{\large SOME OML EXPRESSIONS CONTAINING\\ EQUIVALENCE TERMS}
\label{sec:open}

In \cite{mpoa99} we investigated an equational variety of orthomodular
lattices (OMLs) whose equations hold in the lattice of closed subspaces
of infinite-dimensional Hilbert space ${\cal C}({\cal H})$.  We showed
that this variety could be defined by an infinite set of symmetry
relations for equivalence-like terms.
In the variety we were also able to prove a ``distributivity of
equivalence terms'' in Theorem~7.2 of \cite{mpqo01} (we called it the
``distributivity of identity terms''), shown as Eq.~(\ref{eq:theequation})
below.  In the two papers we conjectured that this ``distributivity''
might hold in every OML, but were missing the proof.  In the meantime we
succeeded in finding one and we provide it below.  We also prove several
related equations that answer a number of open questions in those
papers.  All of these results are primarily a consequence of a more
general result expressed as Eq.~(\ref{eq:theeq1}) below.

We use the notation $a\equiv b$ as an abbreviation for $a\equiv_5 b$.

\begin{theorem}\label{th:theequation}
The following equations hold in all {\em OML}s.
\begin{eqnarray}
\lefteqn{(a \to_1 b) \cap (b \to_2 c)
       \cap (c \to_1 d) \cap (d \to_2 a)  =}\qquad\qquad\nonumber\\
   & & (a \equiv b) \cap (b \equiv c) \cap (c \equiv d)
       \label{eq:theeq1}\\
\lefteqn{(a \to_5 b) \cap (b \to_5 c)
       \cap (c \to_5 d) \cap (d \to_5 a)  =}\qquad\qquad\nonumber\\
   & & (a \equiv b) \cap (b \equiv c) \cap (c \equiv d)
       \label{eq:theeqi5}\\
\lefteqn{(a \to_1 b) \cap (b \to_2 c) \cap (c \to_1 a)
\ \le\  a \equiv c }\qquad\qquad
\label{eq:theeq2} \\
\lefteqn{(a\equiv b)\cap((b\equiv c)\cup(a\equiv c))=}\qquad\qquad\nonumber\\
   & & ((a\equiv b)\cap(b\equiv c))\cup
((a\equiv b)\cap(a\equiv c)) \label{eq:theequation}\\
\lefteqn{(a\equiv b)\cap((b\equiv c)\cup (a\equiv c))\ \le\
a\equiv c}\qquad\qquad \label{eq:om-alt-v}\\
\lefteqn{(a\equiv b)\to_0((a\equiv c)\equiv(b\equiv
    c))\ =\ 1\,.}\qquad\qquad\label{eq:woml2}
\end{eqnarray}
\end{theorem}
\begin{proof}
For Eq.~(\ref{eq:theeq1}), we have
\begin{eqnarray}
\lefteqn{(a \to_1 b) \cap (b \to_2 c)
       \cap (c \to_1 d) \cap (d \to_2 a)}\nonumber\\
&&\qquad= (b \to_2 c)
       \cap (c \to_1 d) \cap (d \to_2 a) \cap (a \to_1 b)\nonumber\\
&&\qquad= ((b' \cap c')
       \cup (c \cap d)) \cap ((d' \cap a') \cup (a \cap b))\nonumber\\
&&\qquad= (b' \cap c' \cap  d' \cap a')
    \cup  (b' \cap c' \cap a \cap b)\nonumber\\
&&\qquad\qquad\qquad
    \cup (c \cap d \cap  d' \cap a')
    \cup  (c \cap d \cap a \cap b)\nonumber\\
&&\qquad= (b' \cap c' \cap  d' \cap a')
    \cup  0
    \cup 0
    \cup  (c \cap d \cap a \cap b)\nonumber\\
&&\qquad= (a \equiv b) \cap (b \equiv c) \cap (c \equiv d)\,.\nonumber
\end{eqnarray}
For the second step we used Lemma 3.14 of \cite{mpoa99}.  For the third
step we used the Marsden-Herman Lemma, given for example as Corollary
3.3 of \cite[p.~259]{beran}.  For the last step we used Lemma 3.11 of
\cite{mpoa99}.

Eq.~(\ref{eq:theeqi5}) follows easily from Eq.~(\ref{eq:theeq1}), noticing
that $a \equiv b \le a\to_5 b\le a\to_1 b,a\to_2 b$.  We twice use
the transitive law $(a
\equiv b) \cap (b \equiv c) \le a \equiv c$, which is Theorem~2.8
of \cite{mpoa99}, in order establish
\begin{eqnarray}
\lefteqn{(a \equiv b) \cap (b \equiv c) \cap (c \equiv d)=}\nonumber\\
&&\qquad(a \equiv b) \cap (b \equiv c) \cap (c \equiv d)
    \cap (d \equiv a) 
\end{eqnarray}
for the purpose of the proof.

Eq.~(\ref{eq:theeq2}) is obtained from Eq.~(\ref{eq:theeq1}) by substituting
$a$ for $d$, then using in the trivial $a\to_2 a=1$ on the left-hand side
and symmetry of equivalence $a\equiv c=c\equiv a$ on the right-hand
side.

In the proof of Eq.~(\ref{eq:theequation}) in Theorem~7.2 of
\cite{mpqo01}, the only use of the (stronger-than-OML) Godowski
equations was to establish Eq.~(\ref{eq:theeq2}) above.  Since we now have
a proof that Eq.~(\ref{eq:theeq2}) holds in all OMLs, it follows that
Eq.~(\ref{eq:theequation}) also holds in all OMLs.

Eqs.~(\ref{eq:om-alt-v}) and (\ref{eq:woml2}) follow from 
Eq.~(\ref{eq:theequation}) by Theorem~2.9 of \cite{mpoa99}.
\end{proof}

Now we address some open questions answered by this theorem.  In
\cite{mpqo01}, we wondered if Eq.~(\ref{eq:theeq2}) above holds in all
OMLs; the answer is affirmative.  In addition, together with
Eq.~(\ref{eq:theequation}) above this result answers all open questions
posed in the paragraph after Theorem~3.16 of \cite{mpoa99}.
Eqs.~(\ref{eq:theequation}), (\ref{eq:om-alt-v}), and (\ref{eq:woml2})
answer the question, posed after Theorem~2.9 of \cite{mpoa99}, of
whether these equations hold in all OMLs.

Eq.~(\ref{eq:theeqi5}) above extends the 3-variable version of it, given
as Eq.~3.21 of Lemma~3.14 of \cite{mpoa99}, to 4 variables.  This in
turn allows us to prove the assertion of Theorem~3.15 of that paper for
$n=4$ (although that assertion still remains an open problem for $n>4$).
It is unknown whether Eq.~(\ref{eq:theeqi5}) holds in all OMLs when
extended to 5 variables.  An extension of Eq.~(\ref{eq:theeqi5}) to 6 (or
more) variables does not hold in all OMLs, because it fails in the OML
of Fig.~2(a) of \cite{mpoa99}.  [We mention that the extension of
Eq.~(\ref{eq:theeqi5}) to any number of variables {\em does} hold in the
lattice ${\cal C}({\cal H})$, since it is a consequence of Theorem~3.12
of \cite{mpoa99}.]

Recall that a WOML (weakly orthomodular lattice) is an OL in which
the following additional condition is satisfied \cite{mpcommp99}:
\begin{eqnarray}
    (a'\cap (a\cup b))\cup b'\cup (a\cap b)&=&1\,.\label{eq:woml}
\end{eqnarray}
In \cite{mpoa99} we asked whether Eqs.~(\ref{eq:theequation}) and
(\ref{eq:woml2}) above hold in all WOMLs.  The next theorem provides the
answer.

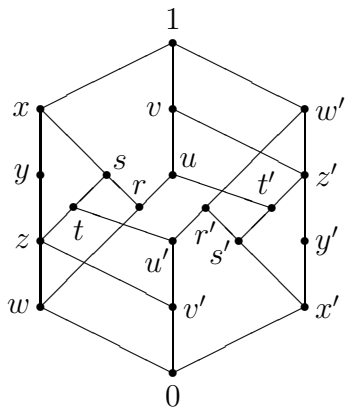
\begin{figure}[htbp]\centering
    \setlength{\unitlength}{1pt}  
    \begin{picture}(260,150)(-10,-10)

      \put(100,0) { 
        \begin{picture}(100,100)(0,0) 
          \put(0,25){\line(0,1){75}}
          \put(0,25){\line(2,-1){50}}
          \put(0,25){\line(1,1){50}}
          \put(0,50){\line(2,-1){50}}
          \put(0,50){\line(1,1){25}}
          \put(0,100){\line(1,-1){37.5}}
          \put(0,100){\line(2,1){50}}
          \put(50,0){\line(0,1){50}}
          \put(50,125){\line(0,-1){50}}
          \put(50,50){\line(-3,1){37.5}}
          \put(100,100){\line(0,-1){75}}
          \put(100,100){\line(-2,1){50}}
          \put(100,100){\line(-1,-1){50}}
          \put(100,75){\line(-2,1){50}}
          \put(100,75){\line(-1,-1){25}}
          \put(100,25){\line(-1,1){37.5}}
          \put(100,25){\line(-2,-1){50}}
          \put(50,75){\line(3,-1){37.5}}

          \put(0,25){\circle*{3}}
          \put(0,50){\circle*{3}}
          \put(0,75){\circle*{3}}
          \put(0,100){\circle*{3}}
          \put(12.5,62.5){\circle*{3}}
          \put(25,75){\circle*{3}}
          \put(37.5,62.5){\circle*{3}}
          \put(50,0){\circle*{3}}
          \put(50,25){\circle*{3}}
          \put(50,50){\circle*{3}}
          \put(50,75){\circle*{3}}
          \put(50,100){\circle*{3}}
          \put(50,125){\circle*{3}}
          \put(62.5,62.5){\circle*{3}}
          \put(75,50){\circle*{3}}
          \put(87.5,62.5){\circle*{3}}
          \put(100,25){\circle*{3}}
          \put(100,50){\circle*{3}}
          \put(100,75){\circle*{3}}
          \put(100,100){\circle*{3}}

          \put(0,25){\makebox(0,0)[r]{$w\ $}}
          \put(0,50){\makebox(0,0)[r]{$z\ $}}
          \put(0,75){\makebox(0,0)[r]{$y\ $}}
          \put(0,100){\makebox(0,0)[r]{$x\ $}}
          \put(12.5,57.5){\makebox(0,0)[t]{$\ t$}}
          \put(28,77){\makebox(0,0)[b]{$\ s$}}
          \put(37.5,67.5){\makebox(0,0)[b]{$r$}}
          \put(50,-5){\makebox(0,0)[t]{$0$}}
          \put(50,25){\makebox(0,0)[l]{$\ v'$}}
          \put(50,47){\makebox(0,0)[t]{$u'\ \ \ $}}
          \put(50,78){\makebox(0,0)[b]{$\ \ \ u$}}
          \put(50,100){\makebox(0,0)[r]{$v\ $}}
          \put(50,130){\makebox(0,0)[b]{$1$}}
          \put(62.5,57.5){\makebox(0,0)[t]{$r'$}}
          \put(70,50){\makebox(0,0)[t]{$s'\ $}}
          \put(87.5,67.5){\makebox(0,0)[b]{$t'\ $}}
          \put(100,25){\makebox(0,0)[l]{$\ x'$}}
          \put(100,50){\makebox(0,0)[l]{$\ y'$}}
          \put(100,75){\makebox(0,0)[l]{$\ z'$}}
          \put(100,100){\makebox(0,0)[l]{$\ w'$}}

        \end{picture}
      }

    \end{picture}

\caption{WOML that violates Eq.~(\ref{eq:theequation}).  [Found by Mike
Rose and Kristin Wilkinson at Argonne National Laboratory with the
program SEM \protect\cite{sem}.] \label{fig:mccune4}}

\end{figure}

\begin{theorem}\label{th:theequationWOML}
Eq.~(\ref{eq:woml2}) holds in all {\rm WOML}s.
Eq.~(\ref{eq:theequation}) does not hold in all {\rm WOML}s.
\end{theorem}
\begin{proof}
Eq.~(\ref{eq:woml2}) holds in all OMLs by Theorem~\ref{th:theequation}.
Since the left-hand-side of Eq.~(\ref{eq:woml2}) evaluates to $1$,
it therefore also holds in all WOMLs by Lemma~3.7 of \cite{mpcommp99}.

Eq.~(\ref{eq:theequation}) fails in the WOML of Figure \ref{fig:mccune4}.
\end{proof}

\bigskip
\section{\large EQUIVALENCES VS.~DIFFERENCES}
\label{sec:s-diff}

A recent paper \cite{dorfer} ``deal[s] with the following
question: What is the proper way to introduce symmetric
differences in orthomodular lattices? Imposing two natural
conditions on this operation, six possibilities remain.''
In this section we show that these ``six possibilities''
are complements of the six equivalence operations
{}from \cite{mpcommp99} and Sec.~\ref{sec:equiv-alg}.
We also draw the reader's attention to the fact that
the Navara's technique of handling two variable OML
expressions used in \cite{dorfer} have previously been
given a computer program support \cite{mpqo01} which
directly gives all needed results. In the end we comment 
on congruence relations from \cite{dorfer} and from 
\cite{mpcommp99}.

Below, on the left-hand sides of the equations are
symmetric differences from Theorem 2 of \cite{dorfer}.
\begin{eqnarray}
a\bigtriangledown b&=&(a\equiv_0b)'\ =\ {\rm B}_9(a,b)\nonumber\\
a\bigtriangleup b&=&(a\equiv_5b)'\ =\ {\rm B}_{84}(a,b)\nonumber\\
a+_l b&=&(a\equiv_1b)'\ =\ {\rm B}_{25}(a,b)\nonumber\\
a+_r b&=&(a\equiv_4b)'\ =\ {\rm B}_{41}(a,b)\nonumber\\
a+_{l'} b&=&(a\equiv_3b)'\ =\ {\rm B}_{73}(a,b)\nonumber\\
a+_{r'} b&=&(a\equiv_2b)'\ =\ {\rm B}_{57}(a,b)\nonumber
\end{eqnarray}

Hence, Definition 1, Theorem 2, most two variable parts
of Propositions 3-14 and of Corollaries 5-13 of
\cite{dorfer} directly follow from \cite{mpcommp99}
(see Sec.~\ref{sec:equiv-alg}) and \cite{mpqo01}
(program \tt beran\rm). E.g., for the proof of Corollary
8 of \cite{dorfer} ($(x+_l y)+_l y=x$) we write:

{\tt beran "-(-(x$\equiv_1$y)$\equiv_1$y)"}

\noindent
and we get the output {\tt 75 x}, where 75 stands for the
Beran expression ($x$).

In \cite{mpcommp99}, it was shown that each of $a\equiv_ib=1$,
$i=0,1,\dots,5$, is a relation of equivalence and of congruence.
Therefore, in an ortholattice, OL or in a weakly orthomodular lattice
(WOML; see Sec.~\ref{sec:open}) there are five such congruence 
relations. In an orthomodular lattice, OML, due to 
Theorem \ref{th:other-eq} they all reduce to the 
following equality: $a=b$. In \cite{dorfer} in Section 3., 
congruence relations are considered in relation to
symmetric differences but are not explicitly defined.
E.g., in Theorem 15 (iii) of  \cite{dorfer} we read:
$a\theta b$ iff $(a\equiv_5b)'\in I$, where I a
a $p$-ideal; in Theorem 15 (iii') we are offered $a\theta b$ iff
$(a\equiv_ib)'\in I$. It would be interesting to know examples  
of $a\theta b$ in orthomodular lattices and which $a\theta b$ 
relations would satisfy the conditions from the afore-mentioned 
Dorfer's Theorem 15 in any orthomodular lattice.   

\bigskip
\section{\large CONCLUSION}
\label{sec:concl}

In Section \ref{sec:equiv-alg} we show that six operations
of equivalence in an orthomodular lattice and the Boolean algebra,
we introduced in \cite{mphpa98}, cannot build equivalence algebras
which would yield orthomodular lattices in the way the implication
algebras.

In Section \ref{sec:open} we show that the distributivity of equivalence
terms holds in any orthomodular lattice which has been an open problem
so far.  We actually prove a more general result, in the form of
Eq.~(\ref{eq:theeq1}), that has as a consequence this distributivity as
well as the answer to several other open problems raised in previous
papers.

In Section \ref{sec:s-diff} we show that six symmetric differences
{}from \cite{dorfer} are nothing but complements of the six
equivalence operations from \cite{mpcommp99}.
We also draw the reader's attention to the fact that
the Navara's technique of handling two variable OML
expressions used in \cite{dorfer} have previously been
given a computer program support \cite{mpqo01} which
directly gives all needed results. In the end we
consider congruence relations from \cite{dorfer} and 
\cite{mpcommp99}.

All two variable expressions used in the paper have been
given their Beran meaning and numbers.

\vfill\eject

\bigskip\bigskip
\parindent=0pt
{\large\bf ACKNOWLEDGEMENTS}

\parindent=20pt
\bigskip

M.~P.~acknowledges supports of the Ministry of Science of Croatia
through the project 0082222.

\vfill\eject

\medskip

\end{document}